\documentclass[twocolumn,,showpacs,preprintnumbers,tbtags]{revtex4}
\usepackage{amsmath}
\usepackage{amssymb}
\usepackage{graphics}
\usepackage{color}
\usepackage{feynmp}
\begin{document}

\let\displaymath=\[
\let\enddisplaymath=\]
\def\[{\begin{equation}}
\def\]{\end{equation}}
\def\<{\begin{myeqnarray}}
\def\>{\end{myeqnarray}}


\makeatletter
\let\old@makecaption=\@makecaption
\def\@makecaption{\small\old@makecaption}
\makeatother

\newcommand{\hypref}[2]{\ifx\href\asklfhas #2\else\href{#1}{#2}\fi}
\newcommand{\remark}[1]{\textbf{#1}}
\newcommand{\figref}[1]{fig. \ref{#1}}
\newcommand{\Figref}[1]{Fig. \ref{#1}}
\newcommand{\tabref}[1]{tab. \ref{#1}}
\newcommand{\indup}[1]{_{\mathrm{#1}}}
\newcommand{\tsum}{{\textstyle\sum}}
\newcommand{\tprod}{{\textstyle\prod}}
\newcommand{\bu}{$\bullet$ \,}

\newcommand{\sfrac}[2]{{\textstyle\frac{#1}{#2}}}
\newcommand{\half}{\sfrac{1}{2}}
\newcommand{\quarter}{\sfrac{1}{4}}

\newcommand{\Op}{\mathcal{O}}\newcommand{\order}[1]{\mathcal{O}(#1)}
\newcommand{\eps}{\varepsilon}
\newcommand{\Lagr}{\mathcal{L}}
\newcommand{\superN}{\mathcal{N}}
\newcommand{\gym}{g_{\scriptscriptstyle\mathrm{YM}}}
\newcommand{\gtwo}{g_2}
\newcommand{\Tr}{\mathop{\mathrm{Tr}}}
\renewcommand{\Re}{\mathop{\mathrm{Re}}}
\renewcommand{\Im}{\mathop{\mathrm{Im}}}
\newcommand{\Li}{\mathop{\mathrm{Li}}\nolimits}
\newcommand{\cdott}{\mathord{\cdot}}
\newcommand{\singlet}{{\mathbf{1}}}

\newcommand{\lrbrk}[1]{\left(#1\right)}
\newcommand{\bigbrk}[1]{\bigl(#1\bigr)}
\newcommand{\vev}[1]{\langle#1\rangle}
\newcommand{\normord}[1]{\mathopen{:}#1\mathclose{:}}
\newcommand{\lrvev}[1]{\left\langle#1\right\rangle}
\newcommand{\bigvev}[1]{\bigl\langle#1\bigr\rangle}
\newcommand{\bigcomm}[2]{\big[#1,#2\big]}
\newcommand{\lrabs}[1]{\left|#1\right|}
\newcommand{\abs}[1]{|#1|}

\newcommand{\nn}{\nonumber}
\newcommand{\nln}{\nonumber\\}
\newcommand{\nl}{\nonumber\\&&\mathord{}}
\newcommand{\nle}{\nonumber\\&=&\mathrel{}}
\newcommand{\eq}{\mathrel{}&=&\mathrel{}}
\newenvironment{myeqnarray}{\arraycolsep0pt\begin{eqnarray}}{\end{eqnarray}\ignorespacesafterend}
\newenvironment{myeqnarray*}{\arraycolsep0pt\begin{eqnarray*}}{\end{eqnarray*}\ignorespacesafterend}

\newcommand{\NPB}[3]{{ Nucl.\ Phys.} {\bf B#1} (#2) #3}
\newcommand{\CMP}[3]{{ Commun.\ Math.\ Phys.} {\bf #1} (#2) #3}
\newcommand{\PRD}[3]{{ Phys.\ Rev.} {\bf D#1} (#2) #3}
\newcommand{\PLB}[3]{{ Phys.\ Lett.} {\bf B#1} (#2) #3}
\newcommand{\PRL}[3]{{ Phys.\ Rev. Lett.} {\bf #1} (#2) #3}
\newcommand{\JHEP}[3]{{JHEP} {\bf #1} (#2) #3}
\newcommand{\hepth}[1]{{\tt hep-th/#1}}
\newcommand{\ft}[2]{{\textstyle\frac{#1}{#2}}}
\newcommand{\cO}{{\cal O}}
\newcommand{\cT}{{\cal T}}
\newcommand\Zb{\bar Z}
\newcommand\bZ{\bar Z}
\newcommand\bF{\bar \Phi}
\newcommand\bP{\bar \Psi}
\newcommand\hint{h_{int}}
\newcommand\hep[1]{[#1]}

\newcommand{\form}{\textsc{Form}}
\newcommand{\mathematica}{\textsc{Mathematica}}


\newcommand{\tr}{\operatorname{tr}}
\newcommand{\del}{\partial}
\newcommand{\diff}[2][\,]{\mathrm{d}^{#1\!}#2}
\newcommand{\pint}[2][d]{\int\!\!\frac{\diff[#1]{#2}}{(2\pi)^#1}}

\newcommand{\stpi}[1][1]{\frac{#1}{16\pi^2}}

\newcommand{\C}{\mathcal{C}}
\renewcommand{\L}{\mathcal{L}}
\renewcommand{\O}{\mathcal{O}}

\newcommand{\al}{\alpha}
\newcommand{\be}{\beta}
\newcommand{\ga}{\gamma}
\newcommand{\de}{\delta}
\newcommand{\ep}{\epsilon}
\newcommand{\vep}{\varepsilon}
\newcommand{\ka}{\kappa}
\newcommand{\si}{\sigma}

\newcommand{\De}{\Delta}
\newcommand{\La}{\Lambda}

\newcommand{\uber}[2]{\left\{ \begin{matrix} #1\cr #2 \end{matrix}\right\}}

\preprint{MPP-2007-145\cr HU-EP-07/49}

\title{Absence of gravitational contributions to the running Yang-Mills coupling}
\author{Dietmar Ebert${}^{1,2}$}
\author{Jan Plefka${}^2$}
\author{Andreas Rodigast${}^2$}
\affiliation{
${}^1$Max-Planck-Institut f\"ur Physik (Werner-Heisenberg-Institut)
F\"ohringer Ring 6, D-80805 M\"unchen, Germany\\ 
${}^2${Institut f\"ur Physik,
Humboldt-Universit\"at zu Berlin,
Newtonstra{\ss}e 15, D-12489 Berlin, Germany}\\
{\tt debert@mppmu.mpg.de, plefka,rodigast@physik.hu-berlin.de}}
\begin{abstract}
The question of a modification of the running gauge coupling of (non-) abelian
gauge theories by an incorporation of the quantum gravity contribution has recently 
attracted considerable interest. In this letter we perform an involved diagrammatical 
calculation in the full Einstein-Yang-Mills system both in cut-off and dimensional 
regularization at one loop order. It is found that all gravitational quadratic 
divergencies cancel in cut-off regularization and are trivially absent in dimensional 
regularization so that there is no alteration to asymptotic freedom at high energies. 
The logarithmic divergencies give rise to an extended effective Einstein-Yang-Mills 
Lagrangian with a counterterm of dimension six.  In the pure Yang-Mills sector this 
counterterm can be removed by a nonlinear field redefinition of the gauge potential, 
reproducing a classical result of Deser, Tsao and van Nieuwenhuizen obtained in the 
background field method with dimensional regularization.
\end{abstract}
\pacs{12.10.Kt, 04.60.-m, 11.10.Hi}

\maketitle
\begin{fmffile}{beta-graphs}
Perturbatively quantized general relativity is well known to be a non-renormalizable
theory due to the negative mass dimension of its 
coupling constant $\kappa$ \cite{PertQG}. The coupling to matter fields 
does generically not improve the situation \cite{PertQG,MatterC,DvN2}, although the possibility
of perturbative finiteness of the maximally supersymmetric gravity theory is still open~\cite{N8finite}.  
Hence, Einstein's theory of gravity does not constitute a fundamental theory as 
its non-renormalizability necessitates 
the inclusion of an infinite set
of higher dimension counterterms in the perturbative quantization process.
Nevertheless, as 
advocated by Donoghue \cite{Donoghue},
it may be treated 
as an approximation to a fundamental theory of quantum gravity 
by the methodology of effective field theories
in order to describe interactions
at scales well below the Planck mass $M_P\sim1/\kappa\sim 10^{19}$GeV. 

In an interesting recent paper Robinson and Wilczek \cite{RobWil} reported on a 
calculation in Einstein gravity coupled to
Yang-Mills theory, where the quantum gravity contributions to the running of the Yang-Mills
coupling $g$ were 
investigated at one-loop order. Interestingly a negative gravitational contribution to the 
Callan-Symanzik $\beta$ function, which quantifies the flow of the Yang-Mills coupling with the
energy scale $E$, was found
\begin{equation}
\label{betag}
\beta_g \equiv\frac{dg}{d \ln E}=-\frac{b_0}{(4\pi)^2}\, g^3 + \frac{a_0}{(4\pi)^2}\, E^2\, \kappa^2\, g\, ,
\end{equation}
with $a_0=-3/2$ in our conventions for the gravitational coupling $\kappa^2$, see
(\ref{EYM}). Irrespective of the non-gravitational value of $b_0$ this would render 
{\sl any} gauge theory assymptotically free 
at energies $E$ close to the Planck mass (including e.g.~pure $U(1)$ Maxwell theory).
Responsible for this effect were quadratic divergencies in one-loop graphs 
containing the graviton propagator \cite{RobPhD}.
As a number of discussed scenarios of physics 
beyond the standard model contain higher dimensional gravity theories
with small gravitational scales, 
such an effect might be observable at the Large Hadron
Collider, as was qualitatively studied in \cite{Followup}. 

However, two recent works have cast doubts on the
results of \cite{RobWil} by reconsidering this effect in Einstein-Maxwell theory. 
The authors of \cite{RobWil} used the background field method, which contains
subtle dependencies on gauge conditions leading to gauge dependent results of the effective action.
To our undestanding the essence of the problem in \cite{RobWil} lies in expanding the gravitational 
field about
flat Minkowski space and the gauge field about a nontrivial solution of the {\sl flat} space 
Yang-Mills field equations, which does {\sl not} constitute a solution of the coupled Einstein-Yang-Mills system.
Indeed in \cite{Pietrykowski} Pietrykowski repeated the analysis of \cite{RobWil}
in the abelian theory employing a parameter
dependent gauge condition and obtained a gauge dependent result.
In a very recent analysis Toms \cite{Toms} pointed out the aforementioned subtleties in the
background field approach and studied the problem in Einstein-Maxwell theory employing
an elaborated 
gauge invariant and gauge condition independent formulation of the background field method 
due to 
Vilkovisky and De Witt \cite{vdW}.
Here dimensional regularization was used and a
{\sl vanishing} gravitational contribution to the gauge coupling $\beta$--function was found. 

In view of these discussions it should be stressed that the question of one loop divergencies in the 
Einstein-Yang-Mills theory was already studied more than 30 years ago in the classical work of 
Deser, Tsao and van Nieuwenhuizen \cite{DvN2}. In a background field approach considering fluctuations about a general
gravitational and gauge field background the necessary one loop counterterms were computed by these authors in 
dimensional regularization. In the pure Yang-Mills sector only one dimension--six counterterm of the
form $\Tr[(D_\mu F^{\mu\nu})^2]$ arose, which moreover must be removed through a nonlinear field 
redefinition of the gauge field.
Hence already these 
results were in contradiction with the findings of \cite{RobWil}.
However, one might be worried about the use of dimensional regularization in all these computations, it
being insensitive to quadratic divergencies which at the same time give rise to a nonvanishing 
$a_0$ in the Callan-Symanzik $\beta$--function (\ref{betag}) as reported in \cite{RobWil} using a cutoff
prescription.

\begin{figure*}[t]
\centering
\begin{tabular}{rc@{\qquad}rc@{\qquad}rc@{\qquad}l}
\raisebox{4ex}{(a)}&
\begin{fmfgraph*}(3,3)
\fmfleft{i}
\fmfright{o}
\fmf{boson}{i,v1}
\fmf{dbl_wiggly,left,tension=0.3}{v1,v2}
\fmf{boson,right,tension=0.3}{v1,v2}
\fmf{boson}{v2,o}
\fmfdot{v1,v2}
\end{fmfgraph*}&
\raisebox{4ex}{(b)}&
\begin{fmfgraph*}(3,3)\fmfkeep{grav_prop_tad}
\fmfleft{i}
\fmfright{o}
\fmftop{t}
\fmf{boson}{i,v1}
\fmf{boson}{v1,o}
\fmffreeze
\fmf{dbl_wiggly,left,tension=0.3}{v1,t,v1}
\fmfdot{v1}
\end{fmfgraph*}&&
&\raisebox{4ex}{$\sim\kappa^2$}\\
\raisebox{4ex}{(c)}&
\begin{fmfgraph*}(3,3)
\fmftop{t}
\fmfbottom{i,o}
\fmf{boson,tension=1.5}{t,v2}
\fmf{boson,tension=1.5}{i,v1}
\fmf{boson,tension=1.5}{v3,o}
\fmf{boson}{v1,v2}
\fmf{boson}{v2,v3}
\fmf{dbl_wiggly,tension=0}{v1,v3}
\fmfdot{v1,v2,v3}
\end{fmfgraph*}&
\raisebox{4ex}{(d)}&\
\begin{fmfgraph*}(3,3)
\fmftop{t}
\fmfbottom{i,o}
\fmf{boson}{t,v2}
\fmf{phantom}{i,v2}
\fmf{boson}{v2,o}
\fmffreeze
\fmf{boson,tension=1}{i,v1}
\fmf{boson,tension=0.4}{v1,v2}
\fmf{dbl_wiggly,tension=0.4,left}{v1,v2}
\fmfdot{v1,v2}
\end{fmfgraph*}&
\raisebox{4ex}{(e)}&
\begin{fmfgraph*}(3,3)
\fmftop{t}
\fmfbottom{i,o}
\fmf{boson}{t,v}
\fmf{boson}{i,v}
\fmf{boson}{v,o}
\fmf{dbl_wiggly,tension=0.6,left}{v,v}
\fmfdot{v}
\end{fmfgraph*}&\raisebox{4ex}{$\sim g\kappa^2$}\\
\end{tabular} 
\caption{\label{fig:grloopcorr}Graviton loop corrections to the gluon two and three point functions.}
\end{figure*}

In view of the recent interest in the problem, its potential experimental 
relevance and the discussed technical 
controveries we have conducted a conceptually straightforward but
involved diagrammatical calculation 
in the full Einstein-Yang-Mills system in cut-off {\sl and} 
dimensional regularization. For this the
gravitational contributions to the gluon self-energy and vertex function 
were evaluated at one loop order. 
The relevant diagrams are listed in figure \ref{fig:grloopcorr}. We find that all quadratic divergencies 
cancel in cut-off regularization, 
and they are trivially absent in dimensional regularization.
The logarithmic divergent contributions are substracted by precisely the
dimension--six 
countertem found by Deser, Tsao and van Nieuwenhuizen \cite{DvN2} confirming their results 
and showing explicitly the non-renormalizability of the Einstein-Yang-Mills theory. 
In particular the $\beta$--function of the Yang-Mills coupling constant $g$
receives {\sl no} gravitational contributions at the one loop order, in contradiction to the
findings of \cite{RobWil}, i.e.~$a_0=0$ in (\ref{betag}). 

The starting point of our analysis is the Einstein-Yang-Mills theory
\begin{equation}
\label{EYM}
\L_\text{EYM}= \frac{2}{\ka^2}\sqrt{-\mathbf g}\, \mathbf{R} 
-\tfrac{1}{2}\sqrt{-\mathbf g}\:\mathbf{g}^{\mu\rho}\mathbf{g}^{\nu\sigma}\Tr\left[ F_{\mu\nu}F_{\rho\sigma} \right] 
\, ,
\end{equation}
with the Ricci scalar $\mathbf{R}$ and the Yang-Mills field strength $F_{\mu\nu}=\partial_\mu A_\nu -
\partial_\nu A_\mu - i\, g\, [A_\mu,A_\nu]$. In the renormalization process we will be led to add 
additional dimension--six operators to this theory, as discussed above.
The metric tensor is split into a flat Minkowski background $\eta_{\mu\nu}=\text{diag}(1,-1,-1,-1)$ 
and the graviton field $h_{\mu\nu}$
\begin{equation}
 \mathbf g_{\mu\nu}= \eta_{\mu\nu}+\ka\, h_{\mu\nu}.
\end{equation}
The Einstein-Yang-Mills lagrangian is then expanded up to second order in $h_{\mu\nu}$ in order to read
off the relevant propagator (modulo gauge fixing) and gluon-graviton vertices occuring in figure 
\ref{fig:grloopcorr}. We work in Feynman gauge for the gluons and in harmonic (de Donder) gauge for
the gravitons with the gauge condition
$
\partial^\nu h_{\mu\nu}=\frac{1}{2}\partial_\mu h
$.
The resulting graviton propagator reads
\begin{equation}
\raisebox{-0.4cm}{\begin{fmfgraph*}(3,2)
\fmfleft{i}
\fmflabel{$\al\be$}{i}
\fmfright{o}
\fmflabel{$\ga\de$}{o}
\fmf{dbl_wiggly,label=$p$}{i,o}
\end{fmfgraph*}}\qquad=\frac{i\left(I^{\al\be,\ga\de}-\frac{1}{d-2}\eta^{\al\be}\eta^{\ga\de}\right)}{p^2+i 0}\, .
\end{equation}
with $I^{\mu\nu,\al\be}\equiv\frac{1}{2}(\eta^{\mu\al}\eta^{\nu\be}+\eta^{\mu\be}\eta^{\nu\al})$.
As gravitational ghosts do not couple to gluons they do not appear in our one-loop computations.
We furthermore note the two gluon--one graviton vertex \footnote{The brackets $[\ldots]$ 
and $(\ldots)$ denote unit weight 
anti--symmetrization and symmetrization respectively, i.e. $\Lambda_{[a_1\ldots a_n]}=
\frac{1}{n!}(\Lambda_{a_1\ldots a_n}\pm \text{permutations})$ and $\Lambda_{(a_1\ldots a_n)}=
\frac{1}{n!}(\Lambda_{a_1\ldots a_n}+\text{permutations})$.}
\begin{equation}
\raisebox{-2.0cm}{\begin{fmfgraph*}(3.5,3.5)
\fmfleft{bi}
\fmfv{l=$\mu\:a$,l.a=90}{bi}
\fmfright{bo}
\fmfv{l=$\nu\:b$,l.a=90}{bo}
\fmftop{g}
\fmflabel{$\al\:\be$}{g}
\fmf{boson,label=$p$}{bi,v}
\fmf{boson,label=$q$}{v,bo}
\fmf{dbl_wiggly}{g,v}
\fmfdot{v}
\end{fmfgraph*}} = -\begin{array}[t]{@{}l@{}}i\ka\delta^{ab}\big[ P^{\mu\nu,\al\be}p\!\cdot\! q +\eta^{\mu\nu}p^{(\al} q^{\be)}  \\
				+\tfrac{1}{2}\eta^{\al\be}p^\nu q^\mu-p^\nu\eta^{\mu(\al} q^{\be)}\\
				-q^\mu\eta^{\nu(\al}p^{\be)}\big]
                                                     \end{array}
\end{equation}
and the two gluon--two graviton vertex
\begin{equation}
\raisebox{-1.5cm}{\begin{fmfgraph*}(3.5,3.5)
\fmfleft{bi}
\fmfv{l=$\mu\:a$,l.a=90}{bi}
\fmftop{bo}
\fmflabel{$\nu\:b$}{bo}
\fmfright{gi}
\fmfv{l=$\al\be$,l.a=90}{gi}
\fmfbottom{go}
\fmflabel{$\ga\de$}{go}
\fmf{boson,label=$p$}{bi,v}
\fmf{boson,label=$q$}{v,bo}
\fmf{dbl_wiggly}{gi,v,go}
\fmfdot{v}
\end{fmfgraph*}}
\quad 
\begin{array}[t]{@{}r@{}l@{}}=\dfrac{i}{2}\ka^2 &\delta^{ab}\big[ (p^\nu q^\mu-p\!\cdot\!q\,\eta^{\mu\nu})P^{\al\be,\ga\de}\\
	&+p\!\cdot\!q( 2I^{\mu\nu,\al(\ga}\eta^{\de)\be} +2I^{\mu\nu,\be(\ga}\eta^{\de)\al}\\
	&\;-I^{\mu\nu,\al\be}\eta^{\ga\de} -I^{\mu\nu,\ga\de}\eta^{\al\be}) \\
	&+2 p^{(\al}q^{\be)}P^{\mu\nu,\ga\de} +2 p^{(\ga}q^{\de)}P^{\mu\nu,\al\be} \\
 	&+\big\{2p^\al \eta^{\nu[\mu}\eta^{\be](\ga}q^{\de)} \\
	&\quad+2p^\ga \eta^{\nu[\mu}\eta^{\de](\al}q^{\be)} \\
	&\quad-p^\nu(q^\al P^{\mu\be,\ga\de} +q^\be P^{\al\mu,\ga\de} \\
	&\quad\quad\;+q^\ga P^{\al\be,\mu\de} +q^\de P^{\al\be,\ga\mu}) \big\} \\
	&+\left\{(p,\mu)\leftrightarrow(q,\nu)\right\} \big] \, ,
	\end{array}
\end{equation}
where we have defined
$
P^{\mu\nu,\al\be}\equiv\frac{1}{2}(\eta^{\mu\al}\eta^{\nu\be}+\eta^{\mu\be}\eta^{\nu\al}-\eta^{\mu\nu}\eta^{\al\be}) 
$.
Using these 
expressions
we have computed the divergent pieces of the 
two--gluon 
graphs (a) and (b) of figure \ref{fig:grloopcorr}
with an incoming momentum of $q$ to be
\begin{align}
\label{gg}
(a)&=\stpi[i] \ka^2(q^2\eta^{\mu\nu}-q^\mu q^\nu)\, \de^{ab}\Big[-\frac{3}{2}\uber{\La^2}{0} \nn \\& - \frac{q^2}{6}\,
\uber{\log \La^2}{\frac{2}{\epsilon}}  + \mbox{finite} \Big]\, , \nn \\
(b)&=\stpi[i]\ka^2(q^2\eta^{\mu\nu}-q^\mu q^\nu)\, \de^{ab}\, \frac{3}{2}\uber{\La^2}{0} \, ,
\end{align}
where we quote both the cut off 
($|k^2|<\Lambda^2$) and dimensional ($d=4-\epsilon$) regularized results.
For details of the calculation see \cite{Rodigast}.
Note the cancellation of the quadratic divergence which already leads to the absence of 
gravitational corrections to the $\beta$ function of $g$ in the abelian theory.

For an abelian gauge theory this would be all there is to
do, as there are no cubic gauge field vertices. In order to substract the remaining
divergence in (\ref{gg}) at a given renormalization
point $\mu$
we have to augment the Einstein-Yang-Mills lagrangian (\ref{EYM}) by novel dimension--six countertems.
Taking into account the Bianchi identity there are naively three possible structures 
\begin{align}
\label{O1O2}
{\cal O}_1&\equiv  \Tr[\, (D_\mu F_{\nu\rho})^2\, ]\, , \qquad
{\cal O}_2\equiv  \Tr[\, (D_\mu F^{\mu}{}_\nu)^2\, ]\, , \nn\\
{\cal O}_3&\equiv i\,\Tr[\, F_\mu{}^\nu\,F_\nu{}^\rho\,F_\rho{}^\mu\,] \, .
\end{align}
However, it turns out that they are linearly related up to total derivative terms
\begin{equation}
\label{linrel}
{\cal O}_2= \frac{1}{2}\,{\cal O}_1 -2\, g\, {\cal O}_3 + \text{total derivatives}\, .
\end{equation}
We are thus led to add the terms
$d_1\, {\cal O}_1$ and $d_2\,{\cal O}_2$ to our lagrangian, where $d_{1,2}$ are the corresponding coupling constants.
Note that the term $d_2\,{\cal O}_2$ in the extended effective lagrangian is proportional to the 
lowest order equations of motion $D_\mu F^{\mu\nu}=0$ and
could be removed by a field redefiniton of the gauge field $A_\mu\rightarrow A_\mu -d_2\, D_{\nu}F^{\nu}{}_\mu/2$
up to higher dimension operators. 

The new two gluon vertices of $d_1\, {\cal O}_1$ and $d_2\,{\cal O}_2$ are
\begin{align}
\raisebox{-0.7cm}{
 \begin{fmfgraph*}(3,3)
 \fmfleft{i}
 \fmflabel{$\mu\:a$}{i}
 \fmfright{o}
 \fmflabel{$\nu\:b$}{o}
 \fmf{boson,label=$q$}{i,v}
 \fmf{boson}{v,o}
 \fmfv{label={$\scriptstyle \O_1$},label.dist=0,decoration.shape=circle,decoration.filled=empty}{v}
 \end{fmfgraph*}} \qquad			&=2i\,d_1\, \de^{ab}q^2(q^2\eta^{\mu\nu}-q^\mu q^\nu)\, ,
\label{11}\\
\raisebox{-0.7cm}{
 \begin{fmfgraph*}(3,3)
 \fmfleft{i}
 \fmflabel{$\mu\:a$}{i}
 \fmfright{o}
 \fmflabel{$\nu\:b$}{o}
 \fmf{boson,label=$q$}{i,v}
 \fmf{boson}{v,o}
 \fmfv{label={$\scriptstyle \O_2$},label.dist=0,decoration.shape=circle,decoration.filled=empty}{v}
 \end{fmfgraph*}} \qquad			&=i\,d_2\, \de^{ab}q^2(q^2\eta^{\mu\nu}-q^\mu q^\nu) \, .
\label{12}
\end{align}
The associated counterterms to these vertices can be used to cancel the logarithmic poles 
of $(a)+(b)$ in (\ref{gg}). However, due to the identical tensor structure of (\ref{11}) and
(\ref{12}) these two counterterms are only fixed up to one free parameter.

In order to fix this free parameter we 
move on to the three--gluon graphs (c), (d) and (e) which probe the non--abelian sector of the theory.
Expanding $\L_\text{EYM}$ to cubic order in gluons and up to 
quadratic 
order in gravitons we read off
the relevant three gluon--one graviton vertex
\begin{equation}
\raisebox{-1.6cm}{\fmfframe(0,1)(1,1){\begin{fmfgraph*}(3.5,3.5)
\fmftop{b1}
\fmfv{l=$\mu\:a$}{b1}
\fmfleft{b2}
\fmfv{l=$\nu\:b$,l.a=90}{b2}
\fmfright{b3}
\fmfv{l=$\rho\:c$,l.a=90}{b3}
\fmfbottom{g}
\fmflabel{$\al\be$}{g}
\fmf{boson,label=$p$}{b1,v}
\fmf{boson,label=$q$,l.s=left}{v,b2}
\fmf{boson,label=$k$,l.s=left}{b3,v}
\fmf{dbl_wiggly}{v,g}
\fmfdot{v}
\end{fmfgraph*}}}
= -g\ka \begin{array}[t]{@{}l@{}}f^{abc}\Big[P^{\al\be,\mu\nu}(p-q)^\rho \\
 	+\eta^{\mu\nu}\eta^{\rho(\al}(p-q)^{\be)} \\
	+\text{cycl}(\mu,p;\nu,q;\rho,k) \Big]\end{array}
\end{equation}
and the involved three gluon-two graviton vertex
\begin{equation}
\raisebox{-2cm}{\fmfframe(0,1)(1,1){\begin{fmfgraph*}(4,4)
\fmfsurround{g1,b1,b2,b3,g2}
\fmflabel{$\mu\:a$}{b1}
\fmflabel{$\nu\:b$}{b2}
\fmflabel{$\rho\:c$}{b3}
\fmflabel{$\al\be$}{g1}
\fmflabel{$\ga\de$}{g2}
\fmf{boson,label=$p$}{b1,v}
\fmf{boson,label=$q$}{b2,v}
\fmf{boson,label=$k$}{b3,v}
\fmf{dbl_wiggly}{g1,v}
\fmf{dbl_wiggly}{g2,v}
\fmfdot{v}
\end{fmfgraph*}}}=\dfrac{1}{2}g\begin{array}[t]{@{}l@{}}\ka^2 f^{abc}\big[(p-q)^\rho(2I^{\mu\nu,\al(\ga}\eta^{\de)\be} \\
		\quad+2I^{\mu\nu,\be(\ga}\eta^{\de)\al} -I^{\mu\nu,\al\be}\eta^{\ga\de} \\
		\quad-I^{\mu\nu,\ga\de}\eta^{\al\be} -\eta^{\mu\nu}P^{\al\be,\ga\de}) \\
		+2\big\{(\eta^{\mu\nu}P^{\ga\de,\rho(\be} \\
		\qquad+I^{\mu\nu,\ga\de}\eta^{\rho(\be})(p-q)^{\al)}\big\} \\
		+\left\{(\al,\be)\leftrightarrow(\ga,\de)\right\} \\
		+\text{cycl}(\mu,p;\nu,q;\rho,k)\big]\, .\end{array}
\label{eq:2g3b}
\end{equation}
With these the evaluation of the 
three--gluon 
graphs of figure 1 can be performed, 
and 
we find the divergent contributions in cut-off and dimensional regularization
\begin{align}
 (c)  =& \stpi g \ka^2 f^{abc} \Bigg\{(\eta^{\mu\nu}(p^\rho(\frac{5}{6}p\!\cdot\! q+\frac{1}{4}q\!\cdot\! k)  \nn\\
	&\qquad\qquad\qquad\quad\;-q^\rho(\frac{5}{6}q\!\cdot\! p+\frac{1}{4}p\!\cdot\! k)) +\dots) \nn\\
	&-\frac{5}{6}( k^\mu k^\nu (p-q)^\rho+\dots) \nn\\
	&-\frac{1}{4}(p^\rho q^\mu k^\nu - p^\nu q^\rho k^\mu)\Bigg\}\uber{\log \La^2}{\frac{2}{\ep}} \, , \label{c}
\\ \nn
(d) =& 
\stpi g \ka^2 f^{abc} \Bigg\{\Big[(\eta^{\mu\nu}(p^\rho(-\frac{7}{6}p\!\cdot\! q-\frac{1}{6}p\!\cdot\! k-\frac{3}{4}q\!\cdot\! k) \nn \\
		&\qquad\qquad\qquad-q^\rho(-\frac{7}{6}q\!\cdot\! p-\frac{1}{6}q\!\cdot\! k-\frac{3}{4}p\!\cdot\! k) +\dots) \nn \\ 
		& +( k^\mu k^\nu (p-q)^\rho+\dots) \nn \\
		&+\frac{3}{4}(p^\rho q^\mu k^\nu - p^\nu q^\rho k^\mu)\Big]\uber{\log\La^2}{\frac{2}{\ep}} \nn\\
&		+\frac{3}{2}(\eta^{\mu\nu}(p-q)^\rho+\dots)\uber{\La^2}{0}\Bigg\} \, ,
\\
(e)  =& -\stpi g \ka^2 f^{abc} \frac{3}{2}(\eta^{\mu\nu}(p-q)^\rho+\dots)\uber{\La^2}{0} \, , \label{e}
\end{align}
where the external gluons carry the labels $(\mu,p;\nu,q;\rho,k)$ and the dots refer to a symmetrization
in these labelings.
Summing these contributions up we again observe a 
cancellation 
of the quadratic divergencies arising
in cut--off regularization. In order to 
subtract 
the remaining 
logarithmic divergencies in the 
three--gluon amplitudes we need to consider the three point vertices of 
$d_1\,{\cal O}_1$ and $d_2\,{\cal O}_2$ emerging from  
(\ref{O1O2})
\begin{equation}
\begin{split}
\raisebox{-1.5cm}[0.7cm][0cm]{
 \fmfframe(1,1)(1,1){\begin{fmfgraph*}(3,3)
 \fmfsurround{p,q,k}
 \fmflabel{$\mu\:a$}{p}
 \fmflabel{$\nu\:b$}{q}
 \fmflabel{$\rho\:c$}{k}
 \fmf{boson,label=$p$}{p,v}
 \fmf{boson,label=$q$}{q,v}
 \fmf{boson,label=$k$}{k,v}
 \fmfv{label={$\scriptstyle \O_1$},label.dist=0,decoration.shape=circle,decoration.filled=empty}{v}
 \end{fmfgraph*}}} = d_1g&f^{abc}\big[\eta^{\mu\nu}(p^\rho(4p \!\cdot\! q +2p\!\cdot\! k) \\
                          &-q^\rho(4q \!\cdot\! p +2q\!\cdot\! k))+\dots\\
 			- &2\left(k^\mu k^\nu (p-q)^\rho +\dots\right) \big] \, ,
\end{split}
\end{equation}
\begin{equation}
\raisebox{-2cm}[0cm][0cm]{
\fmfframe(1,1)(1,1){\begin{fmfgraph*}(3,3)
 \fmfsurround{p,q,k}
 \fmflabel{$\mu\:a$}{p}
 \fmflabel{$\nu\:b$}{q}
 \fmflabel{$\rho\:c$}{k}
 \fmf{boson,label=$p$}{p,v}
 \fmf{boson,label=$q$}{q,v}
 \fmf{boson,label=$k$}{k,v}
 \fmfv{label={$\scriptstyle \O_2$},label.dist=0,decoration.shape=circle,decoration.filled=empty}{v}
 \end{fmfgraph*}}} \begin{array}[t]{@{}r@{}l@{}}
 			=d_2g&f^{abc}\big[\eta^{\mu\nu}(p^\rho(2p\cdot q +p\cdot k +3 q\cdot k)\\
					&\;-q^\rho(2q\cdot p +q\cdot k +3 p\cdot k))+\dots \\
 				-&\left( k^\mu k^\nu (p-q)^\rho +\dots\right) \\
 				-&3(p^\rho q^\mu k^\nu-p^\nu q^\rho k^\mu) \big] \, .
			\end{array}
\end{equation}
Remarkably it turns out that ${\cal O}_2$ alone provides the right tensor structures to remove 
all the divergences of (\ref{c})-(\ref{e}).
Renormalization of the two and three point gluon functions is then performed by 
considering the 
extended effective
Einstein-Yang-Mills theory augmented by
$d_2\,{\cal O}_2$  plus all the associated counter terms.
The complete 
effective lagrangian at the one-loop level
is then of the schematic form
\begin{align}
\L =& \L_{\text{ext}} + \L_{\text{CT}} \, ,\nn\\
\L_{\text{ext}}=& \L_{\text{EYM, ren}} + d_2\, \Tr[(D_{\mu}F^{\mu\nu})^2] \, , \nn\\
\L_{\text{CT}}=& \delta_2\, (\partial A)^2 + g\, \delta^{3g}_1\, A^2 \partial A +  \delta_1^{2d_2}\, 
(\partial^2 A)^2 \nn\\
&+  g\, \delta_1^{3d_2}\, \partial^2A\, \partial A A +{\cal O}(A^4) \, .
\end{align}
The computed divergencies are then substracted at a renormalization scale 
$\mu$
 through a choice of
renormalization conditions. This leads to the following $\kappa^2$ dependencies of the 
dimension--six counterterms
\begin{align}
\label{delta}
\delta_1^{2d_2}\Bigr |_{{\cal O}(\kappa^2)}= \delta_1^{3d_2}\Bigr|_{{\cal O}(\kappa^2)}	
=&\stpi\frac{1}{6}\, \ka^2
\, \uber{\log(\frac{\Lambda^2}{\mu^2})}{ \frac{2\,\mu^{-\ep}}{\epsilon} } \, ,
\end{align}
where the first equality $\delta_1^{2d_2} |_{{\cal O}(\kappa^2)}=\delta_1^{3d_2} |_{{\cal O}(\kappa^2)}$ 
arises as a necessary
condition on the renormalization of the gluon selfenergy and vertex function
and constitutes an independent consistency check of our results
as required by the  Slavnov-Taylor-Ward identity. Also note that this countertem is
in prescise 
agreement 
to the one found by Deser, Tsao and van Nieuwenhuizen \cite{DvN2} 
with the background field  method.
In particular the Yang-Mills 
vertex counterterm $\delta_1$ and wavefunction renormalization
$\delta_2$ receive {\sl no} contributions at order $\kappa^2$.
The counterterms (\ref{delta}) then lead to the following relations of renormalized $g,d_2$ to 
bare $g_0,d_{2,0}$ couplings at one loop order
\begin{equation}
\frac{g}{g_0}=1+\frac{3}{2}\, \delta_2 -\delta_1^{3g} \, , \qquad  
\frac{d_2}{d_{2,0}}=1+ \delta_2 -\frac{\delta_1^{2d_2}}{d_{2,0}} \, .
\end{equation}
This then provides
the following gravitational contributions to the $\beta$ functions of the gauge
coupling $g$ and the novel coupling $d_2$
\begin{equation}
\beta_g\Bigr|_{{\cal O}(\kappa^2)} = 0 , \qquad
\beta_{d_2}\Bigr|_{{\cal O}(\kappa^2)} =\frac{1}{(4\pi)^2}\, \frac{1}{3}\, \kappa^2 \, .
\end{equation}
However, the renormalization of $d_2$ through gravitational interactions is not of 
particular
relevance as this coupling can be removed through a nonlinear field redefinition
as mentioned above. 

Let us qualitatively extend our discussion to the gluon four point function.
A simple dimensional consideration of loop integrals for the gluon four point function yields a new
operator structure $\kappa^4  \Tr(F^4)$,
which cannot be removed by the above quoted field redefinition. This is in agreement with the 
findings of \cite{DvN2}, where it was shown that such terms appear as 
the square of the energy-momentum tensor, $(T_{\mu\nu})^2$. 

Interesting further extension of this work would be to consider large extra dimension scenarios,
where a cut-off prescription appears mandatory, as well as the coupling to additional matter fields. 
The case of gauged $N$ extended supergravities was considered in
\cite{Christensen:1980ee} in the background field method with dimensional regularization, where a 
vanishing of the beta function for $N>4$ was found.


\vspace{6mm}
\noindent
{\bf Acknowledgments} \\[.2cm]
\noindent
We would like to thank Harald Dorn, Dieter L\"ust, Kelly Stelle and Stefan
Theisen for important discussions. D.E. expresses his gratitude to Dieter
L\"ust and his colleagues for the kind hospitality at the Max-Planck-Institute for Physics 
(Werner-Heisenberg-Institute) and for financial support.
Our computation made use of the symbolic manipulation system FORM \cite{Form}.


\end{fmffile}


\end{document}